\documentclass{webofc}
\usepackage[varg]{txfonts}   
\usepackage{wrapfig}
\usepackage{caption}
\usepackage{cases}
\usepackage{empheq}
\begin{document}
\title{Model of the Phase Transition Mimicking the Pasta Phase in Cold and Dense Quark-Hadron Matter}
%
%
\author{\firstname{Alexander} \lastname{Ayriyan}\inst{1}\fnsep\thanks{\email{ayriyan@jinr.ru}} \and
        \firstname{Hovik} \lastname{Grigorian}\inst{1,2}\fnsep\thanks{\email{hovik.grigorian@gmail.com}}
}
\institute{
           Joint Institute for Nuclear Research, Joliot-Curie 6, 141980 Dubna, Russia
\and
           Yerevan State University, Alek Manyukyan 1, 0025 Yerevan, Republic of Armenia
          }
\abstract{
A simple mixed phase model mimicking so-called "pasta" phases in the quark-hadron phase transition is developed and applied to static neutron stars for the case of DD2 type hadonic and NJL type quark matter models. The influence of the mixed phase on the mass-radius relation of the compact stars is investigated. Model parameters are chosen such that the results are in agreement with the observational constraints for masses and radii of  pulsars.
}
\maketitle

\vspace{1cm}

\section{Introduction and Motivation}
\label{intro}

Usually modeling of possible hadron-quark phase transition is made with the use of the so-called Maxwell construction, where the two phases are assumed to be separated. However due to the finite surface tension effects the mixed phase with non-trivial structure (so-called "pasta" phase) could be thermodynamically preferred~\cite{vos_2003, tatsumi_2003, endo_2005, endo_2006, noda_2013, yasutake_2014}. 

A simple model of phase transition construction mimicking the ``pasta'' phase is proposed. The model is parametrized by the use of additional pressure corresponding to the impact of structural effects in the mixed phase to the critical pressure of Maxwell construction. The construction ignoring the details of the mixed phase structure is simple for calculations and therefore it can be easily used for investigations of the influence of the mixed phase on compact star mechanical characteristics (e.g.~\cite{noda_2013}).

The question of the mixed phase existence is topical in the investigations of the hybrid compact star structure and characteristics, which could be compared to the available observational data~\cite{bejger_2005, alvarez_2015}. In this work an application of this model to the structure and mass-radius relation of the static neutron stars is presented using DD2 type hadonic EoS model~\cite{typel_2016} and NJL model with 8-quark interaction~\cite{benic_2014} for quark matter.

The method could be extended to the finite temperature case along to the whole first oder transition border of the QCD phase diagram and could be used in the simulations of heavy ion collisions as well (e.g.~\cite{bugaev_2015, toneev_2010}). The possibility of such kind of extension is also discussed.


\newpage

\section{The model}
\label{model}

Let us suppose that the hadroic and quark matter phases are given with the Equation of State (EoS) $P_{H}(\mu)$ and $P_{Q}(\mu)$ (the dependence of pressure on chemical potential at $T=0$ case relevant for the NS modeling) correspondingly.

It is assumed that the first oder Maxwell phase transition exists between the given two phases with a critical chemical potential $\mu_{c}$ satisfying the following relation:
\begin{equation}
P_{Q}\left(\mu_{c}\right)=P_{H}\left(\mu_{c}\right) = P_c.
\end{equation}

\begin{figure}[ht!]
\captionsetup{justification=centering}
\centering
\includegraphics[width=0.5\textwidth]{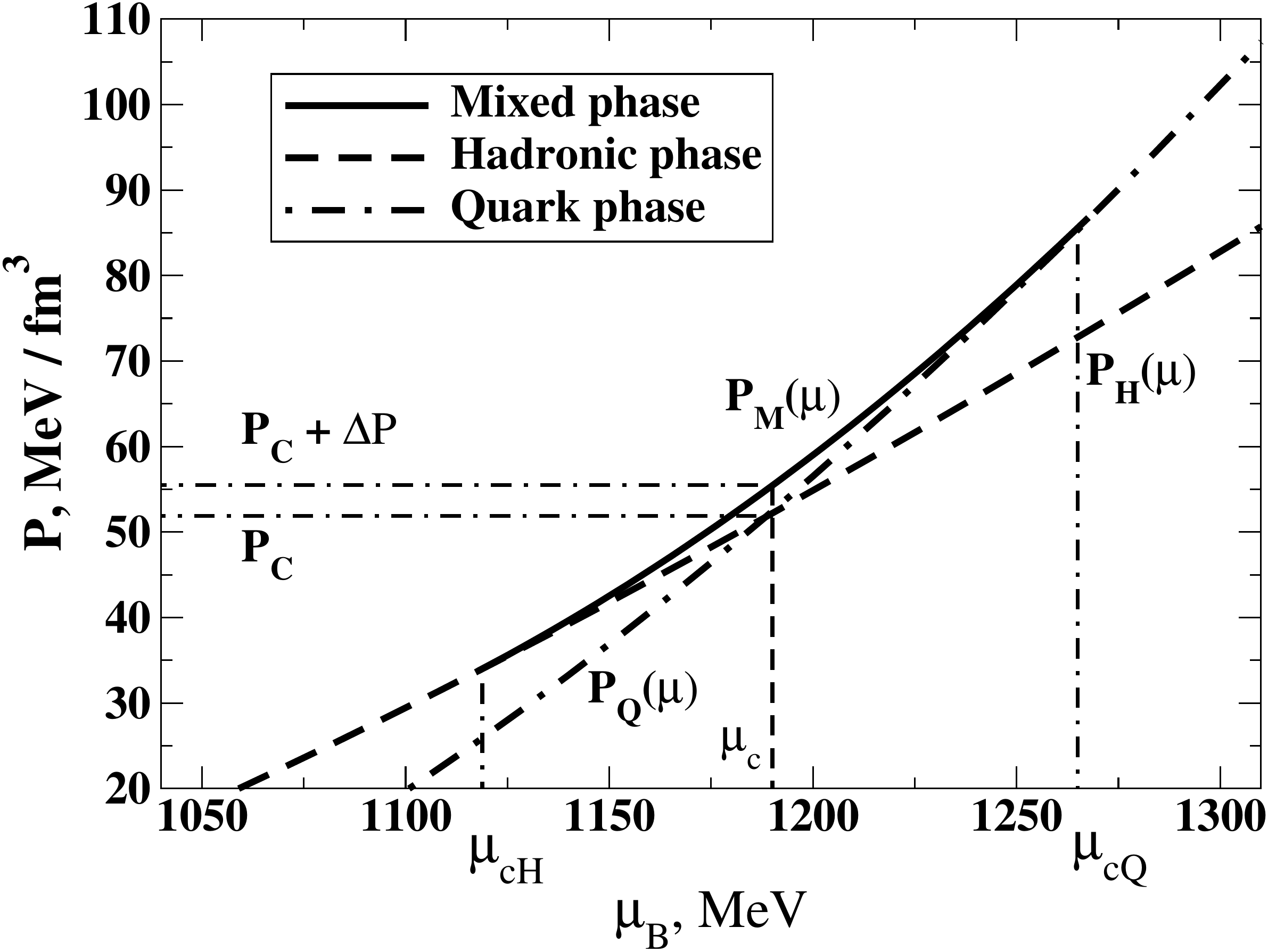}
\caption{A model of the equation of state with mixed phase based on Maxwell construction.}
\label{fig1}
\end{figure}

Now Maxwell construction can be modified with the assumption that close to the phase transition point the Equation of State (EoS) of both phases are changing due to their interactions (see fig.~\ref{fig1}), so that the effective mixed phase EoS can be expressed with the following form 
\begin{equation}
P_{M}(\mu)=\alpha(\mu-\mu_{c})^{p}+\beta(\mu-\mu_{c})^{q}+P_c+\Delta P,
\label{p_mu}
\end{equation}
where $P_c = {P_{H}}_c = {P_{Q}}_c$, while $p>q$ are natural numbers for instance.
Here the additional pressure at $\mu_{c}$ is considered to be the free parameter of the model:
\begin{equation}
P_{M}\left(\mu_{c}\right)=P_{H}\left(\mu_{c}\right)+\Delta P.
\end{equation}
This research is focused on the quadratic form of the mixed phase EoS (e.g. $p=2$ and $q=1$):
\begin{equation}
P_{M}(\mu)=\alpha(\mu-\mu_{c})^{2}+\beta(\mu-\mu_{c})+P_c+\Delta P
\end{equation}
The transition from the $H$-phase to the $M$-phase is smooth without a jump in the density ($n(\mu)=dP(\mu)/d\mu$). Thus there are new critical chemical potentials: $\mu_{cH}$ for the transition from the $H$-phase to the $M$-phase and $\mu_{cQ}$ for the transition from the $M$-phase to the $Q$-phase. So there are four unknowns including the coefficients $\alpha$ and $\beta$ from Eq.~\eqref{p_mu}. The phase transition conditions are

\begin{align}
\begin{cases}
P_{M}(\mu_{cH}) = P_{H}(\mu_{cH}),\\
P_{M}(\mu_{cQ}) = P_{Q}(\mu_{cQ}),
\label{formula:pr}
\end{cases}
\\
\begin{cases}
n_{M}(\mu_{cH}) = n_{H}(\mu_{cH}),\\
n_{M}(\mu_{cQ}) = n_{Q}(\mu_{cQ}).
\label{formula:n_b}
\end{cases}
\end{align}
While $\alpha$ and $\beta$ are expressed from the set of equations for pressure 
\begin{eqnarray}
\begin{cases}
\displaystyle	\alpha  =  \frac{1}{\mu_{cQ}-\mu_{cH}}\left(\frac{P_{Q}-(\Delta P + P_c)}{\mu_{cQ}-\mu_{c}}
	- \frac{P_{H}-(\Delta P + P_c)}{\mu_{cH}-\mu_{c}}\right),\\[5mm]
\displaystyle	\beta  =  \frac{P_{H} - (\Delta P + P_c)}{\mu_{cH}-\mu_{c}}
	+\frac{P_{Q}-(\Delta P + P_c)}{\mu_{cQ}-\mu_{c}}
	-\frac{P_{Q}-P_{H}}{\mu_{cQ}-\mu_{cH}}~,
\end{cases}
\end{eqnarray}
and are eliminated form the set of equations for densities (eq.~\eqref{formula:n_b}). Then solving the remaining equations for the densities one can find the values for the critical chemical potentials $\mu_{cH}$ and $\mu_{cQ}$.

\subsection{Results}
\label{rslts}
\begin{figure}[ht!]
\centering
\captionsetup{justification=centering}
\includegraphics[width=0.5\textwidth]{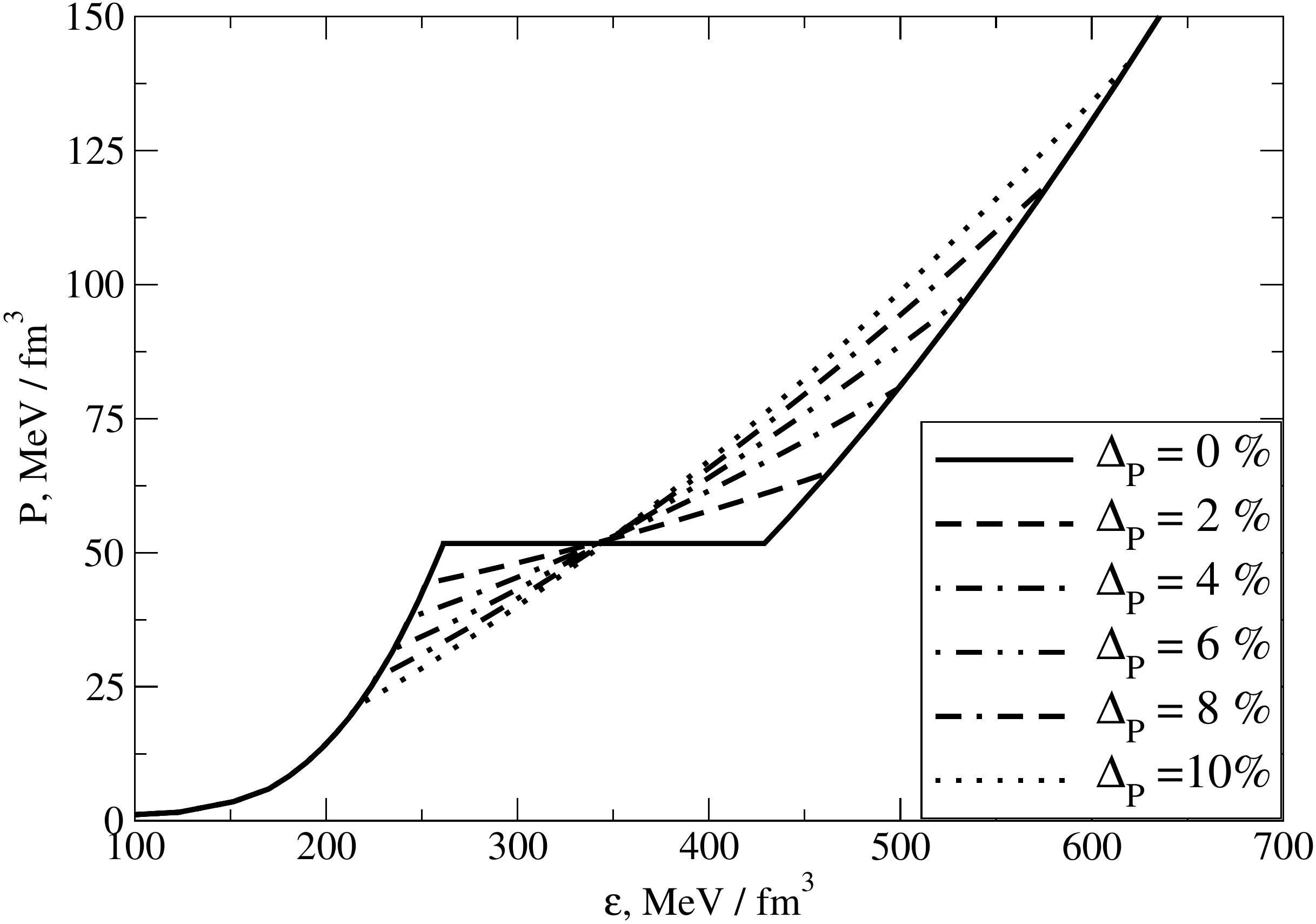}
\caption{Influence of the additional pressure in the mixed phase on the phase transition.}
\label{fig2}
\end{figure}

For the illustration purposes, the described model is applied for the construction the mixed phase EoS  (see~fig.~\ref{fig1}) between hadronic and quark matter respectively given by the DD2 type model~\cite{typel_2016} amd NJL model with 8-quark interaction~\cite{benic_2014}.
The results of the construction are shown in fig.~\ref{fig2}. Here the varied pressure is given by unitless parameter $\Delta_P=\Delta P / P_c$. The parameter $\Delta_P$ is applied up to $10\%$. This completely covers the whole range of the additional pressure obtained by macroscopic models of the mixed phase~\cite{vos_2003, tatsumi_2003, endo_2005, endo_2006, yasutake_2014}. Note, that $P$-$\varepsilon$ dependence of mixed phase by the construction is not a linear function, but is very close~to~it.

With the use of these set of the EoS models the mechanical characteristics of the compact stars are investigated. 

\begin{figure}[ht!]
\centering
\captionsetup{justification=centering}
\includegraphics[width=0.5\textwidth]{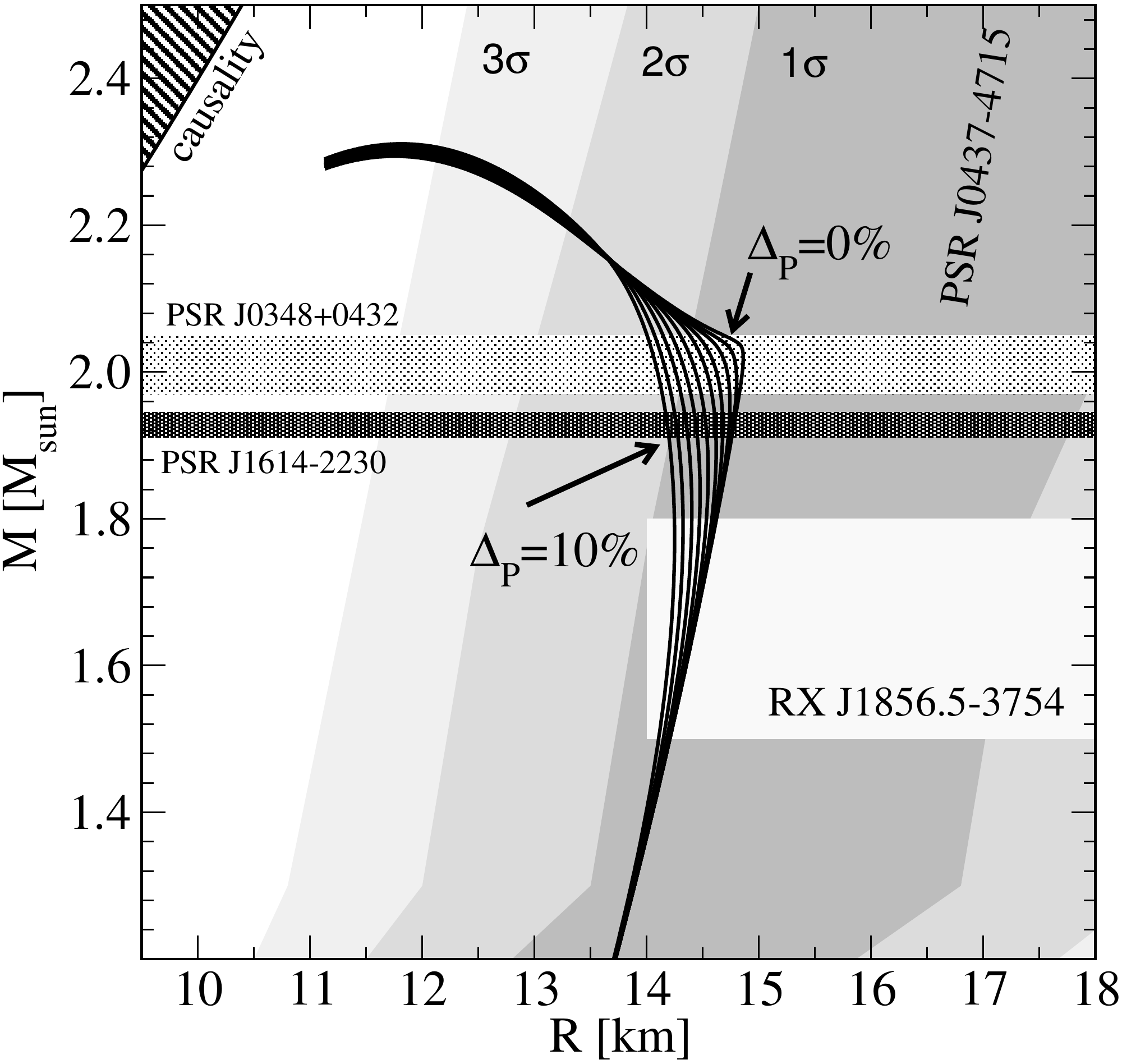}
\caption{Influence of the mixed phase on the mass-radius relation.}
\label{fig3}
\end{figure}

The mass and radius relations of static spherical symmetric neutron stars are given as a solution of the Tolman-Oppenheimer-Volkoff (TOV) equations \cite{tolman_1939, oppen_1939}:
\begin{eqnarray}
\begin{cases}
\displaystyle \frac{dP(r)}{dr} = - \frac{G M( r)\varepsilon( r)}{r^2}\frac{\left(1+\frac{P( r)}{\varepsilon( r)}\right)
\left(1+\frac{4\pi r^3 P( r)}{M( r)}\right)}{\left(1-\frac{2GM( r)}{r}\right)},\\
\displaystyle \frac{dM( r)}{dr} = 4\pi r^2 \varepsilon( r),
\end{cases}
\end{eqnarray}
where $\hbar=c=1$, $r$ is the coordinate distance from the center, $M(r)$, $\varepsilon(r)$ and $P(r)$ are the mass, the energy density and the pressure profiles correspondingly, and $G$ is the gravitational constant.
Each of these profiles, as a solution of TOV equations depends on the central value of the energy density $\varepsilon(r=0)$. The relation between the radius $R$ where the pressure is $P(r=R)=0$ and the total mas $M=M(r=R)$ are shown in the fig.~\ref{fig3} for different values of the parameter $\Delta_P$.


The presence of the mixed phase changes softness of the EoS and the maximum mass becomes larger when the maximum radius decreases. The heaviest known pulsars PSR~J0348$+$0432 and PSR~J1614$-$2230 have respectively the following masses $2.01^{+0.04}_{-0.04}\text{ M}_{\odot}$~\cite{antoniadis_2013} and $1.928^{+0.017}_{-0.017}\text{ M}_{\odot}$~\cite{demorest_2010,fonseca_2016}. Therefore the maximum possible mass given by a model should exceed this limits. These constraints are shown in the fig.~\ref{fig3}. It is demonstrated that all considered EoS models with the mixed phase satisfy this requirement. On this plot the constrains from the radius estimations of two pulsars PSR~J0437$-$4715~\cite{bogdanov_2013} and RX~J1856.5$-$3754~\cite{hambaryan_2014} are also shown and the results are in agreement with this.

\section{Conclusions and Discussion}
\label{conc}

A simple model of hadron-quark phase transition mimicking the "pasta" structure is proposed. It is parametrized using the additional pressure due to the structure of the mixed phase.
The model is based on standard Maxwell construction of the first order phase transitions. It is ignoring the details of the mixed phase structure, at the same time it is easy to use for investigations of the influence of the mixed phase on compact star mechanical characteristics. The structure and mass-radius relations of the static neutron stars are presented for the case of the DD2 type hadonic vs NJL type quark matter EoS models. It is shown that the variation of the parameter in the relevant range of the additional pressure, for the mixed phase, changes the maximum mass and the radius of the star configurations in the limits tolerated by observational constraints.

\begin{figure}[h!]
\centering
\captionsetup{justification=centering}
\includegraphics[width=0.5\textwidth]{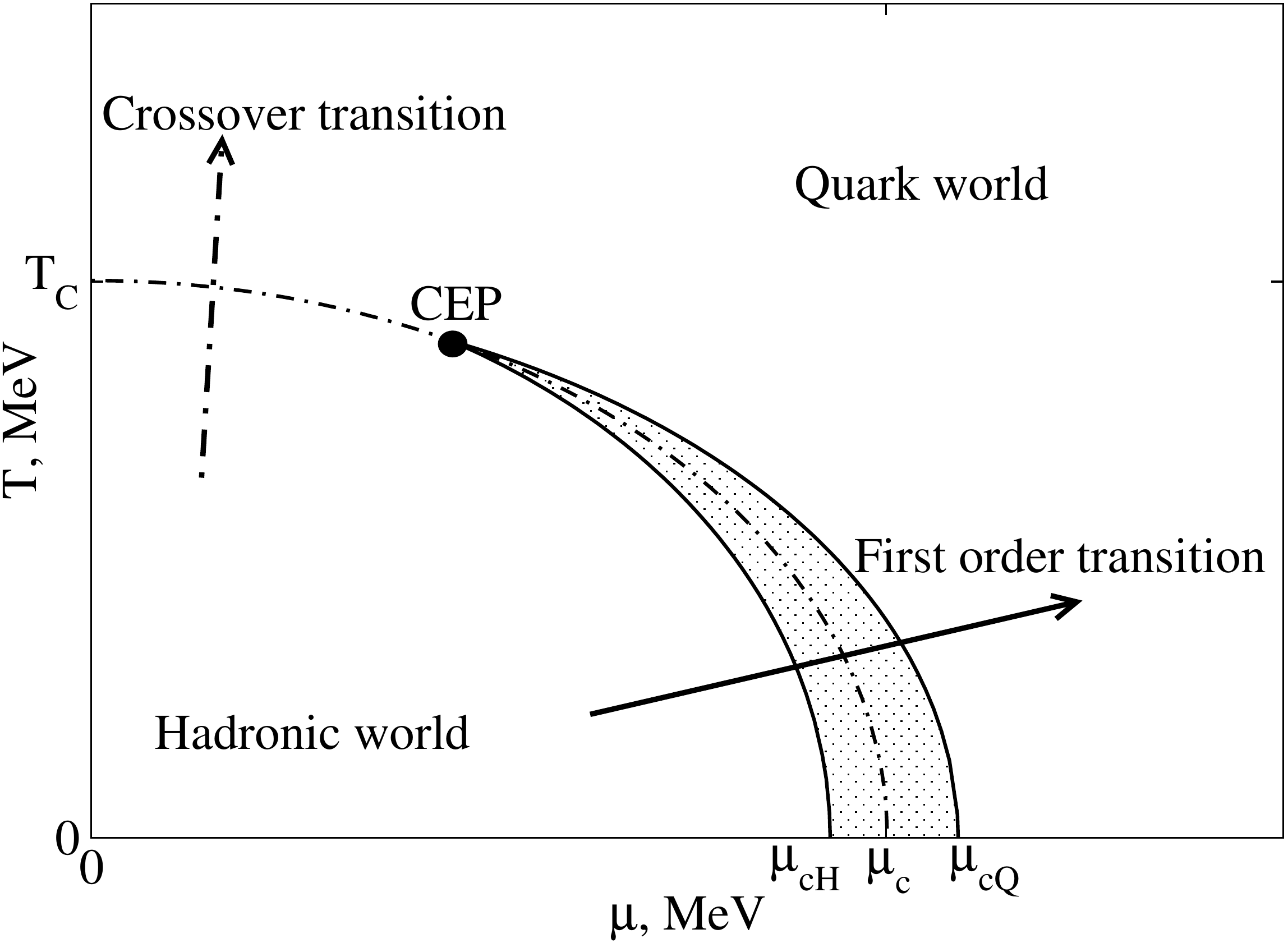}
\caption{A simple schematic view to the QCD phase diagram. The mixed phase of the first order hadron-quark phase transition is colored grey.}
\label{fig_pht}
\end{figure}

The~method could be extended to the finite temperature case along the whole first oder transition border of the QCD phase diagram as it is shown in the fig.\ref{fig_pht} assuming that the additional pressure decreases proportionally to increases of the temperature. Thus the case when the additional pressure goes from the value $\Delta_P$ at $(\mu=\mu_c,T=0)$ down to the $0$ at Critical End Point (CEP) is demonstrated in the fig.~\ref{fig_pht}. So, the first order phase transition (the jump in density) is disappearing at CEP, therefore the possible mixed phase is also disappearing. For the crossover transition one needs to apply another construction. The considered approach for the transition construction is independent of internal details of the transition, therefore it could be applied both for beta-equilibrated and for symmetric nuclear matter. Thus it could be used in the simulations of heavy ion collisions.

\section*{Acknowledgements}
The authors thank Prof.~David Blaschke, Prof.~Dmitry Voskresensky, Dr.~David Edwin Alvarez Castillo, Dr. Vahagn Abgaryan and Konstantin Maslov for fruitful discussions. Prof.~Stefan Typel and Dr. Sanjin Benic provided tables of the EoS models, special thanks to them. The research was carried out under financial support of the Russian Science Foundation (project No. 17-12-01427).
%
%
%

\end{document}